\documentclass[[aps,
prl,
reprint,
superscriptaddress,
groupedaddress,
amsmath,amssymb,
floatfix,
12pt, 
]{revtex4}
\usepackage{color,soul} 
\usepackage{graphicx}
\usepackage{wrapfig}
\usepackage{dcolumn}
\usepackage{bm}
\usepackage{seqsplit} 
\usepackage{ulem} 
\usepackage{float}

\begin{document}


\title{Extreme Near-Field Heat Transfer Between Gold Surfaces}

\author{Takuro Tokunaga}
\affiliation{
Department of Mechanical Engineering, University of Utah, Salt Lake City, Utah 84112, United States.}
 
\author{Amun Jarzembski}
\affiliation{
Sandia National Laboratories, Albuquerque, NM 87185, United States.}

\author{Takuma Shiga}
\affiliation{
Department of Mechanical Engineering, The University of Tokyo, Bunkyo, Tokyo 113-8656, Japan.}

\author{Keunhan Park}
\email{kpark@mech.utah.edu}
\affiliation{
Department of Mechanical Engineering, University of Utah, Salt Lake City, Utah 84112, United States.}

\author{Mathieu Francoeur}
\email{mfrancoeur@mech.utah.edu}
\affiliation{
Department of Mechanical Engineering, University of Utah, Salt Lake City, Utah 84112, United States.}


\begin{abstract}
Extreme near-field heat transfer between metallic surfaces is a subject of debate as the state-of-the-art theory and experiments are in disagreement on the energy carriers driving heat transport. In an effort to elucidate the physics of extreme near-field heat transfer between metallic surfaces, this Letter presents a comprehensive model combining radiation, acoustic phonon and electron transport across sub-10-nm vacuum gaps. The results obtained for gold surfaces show that in the absence of bias voltage, acoustic phonon transport is dominant for vacuum gaps smaller than $\sim$2 nm. The application of a bias voltage significantly affects the dominant energy carriers as it increases the phonon contribution mediated by the long-range Coulomb force and the electron contribution due to a lower potential barrier. For a bias voltage of 0.6 V, acoustic phonon transport becomes dominant at a vacuum gap of 5 nm, whereas electron tunneling dominates at sub-1-nm vacuum gaps. The comparison of the theory against experimental data from the literature suggests that well-controlled measurements between metallic surfaces are needed to quantify the contributions of acoustic phonon and electron as a function of the bias voltage. 
\end{abstract}

\maketitle
Radiative heat transfer between two surfaces separated by a sub-wavelength vacuum gap can exceed the far-field blackbody limit by a few orders of magnitude owing to tunneling of evanescent electromagnetic (EM) waves \cite{Polder1971}. Theoretical predictions of near-field radiative heat transfer based on fluctuational electrodynamics \cite{Rytov1987} are well-established, and have been experimentally validated in various configurations for nanosized vacuum gaps \cite{Rousseau2009,Shen2009,Song2015,Song2016a,St-Gelais2016,Bernardi2016,Watjen2016,Ito2017,Fiorino2018,Lim2018,Ghashami2018a,DeSutter2019,Shi2019,Tang2020,Ying2020}. However, fluctuational electrodynamics may not be able to accurately describe heat transfer in the extreme near-field regime, defined here as sub-10-nm vacuum gap distances, due to its macroscopic nature involving local field averaging (i.e., local dielectric function) \cite{Chiloyan2015}. In addition, fluctuational electrodynamics solely accounts for EM waves (i.e., radiation), whereas acoustic phonons and electrons may also contribute to heat transport between surfaces separated by single-digit nanometer vacuum gaps prior to contact. A few theoretical works have investigated acoustic phonon transport across vacuum gaps \cite{Prunnila2010,Ezzahri2014,Xiong2014,Chiloyan2015,Pendry2016,Sasihithlu2017,Zhang2018,Messina2018,Alkurdi2020}, and some have explicitly shown the inadequacy of fluctuational electrodynamics for modeling heat transfer in the extreme near field \cite{Ezzahri2014,Chiloyan2015,Pendry2016,Sasihithlu2017,Zhang2018,Messina2018,Alkurdi2020}. 

Experimental measurements of extreme near-field heat transfer are scarce \cite{Kittel2005a, Altfeder2010, Kim2015,Kloppstech2017,Cui2017,Jarzembski2019}. Jarzembski {\it{et al}}. \cite{Jarzembski2019} measured a thermal conductance exceeding fluctuational electrodynamics predictions by three orders of magnitude between a silicon tip and a platinum nanoheater separated by sub-10-nm vacuum gaps down to contact. By considering all energy carriers, it was shown quantitatively that heat transfer across vacuum gaps was dominated by acoustic phonon transport mediated by van der Waals and Coulomb forces. Extreme near-field heat transfer between metallic surfaces have also been measured, but drastically different results have been reported. Kim \textit{et al.} \cite{Kim2015} measured heat transfer between a thermocouple-embedded tip fabricated at the free-end of a stiff cantilever and a suspended resistive microheater at 305 K. Their measurements between a gold (Au)-coated tip and an Au surface are in good agreement with fluctuational electrodynamics down to a vacuum gap of $\sim$2 nm. On the other hand, Kloppstech \textit{et al.} \cite{Kloppstech2017} reported the measurement of heat transfer between an Au thermocouple-embedded scanning tunneling microscope tip and a cryogenically-cooled Au surface, observing heat transfer largely exceeding fluctuational electrodynamics for vacuum gaps from 7 nm down to 0.2 nm. However, no comprehensive modeling was performed to quantitatively support their observation.

Subsequently, Cui \textit{et al.} \cite{Cui2017} measured heat transfer between an Au-coated tip and an Au surface, both subjected to various surface-cleaning procedures, for vacuum gaps from 5 nm down to a few {\AA}. They hypothesized that the large heat transfer reported by Kloppstech {\it{et al}}. \cite{Kloppstech2017} may be due to a low apparent potential barrier for electron tunneling mediated by surface contaminants bridging the tip and the surface prior to bulk contact.  

A few theoretical works have analyzed extreme near-field heat transfer between Au surfaces \cite{Pendry2016,Zhang2018,Messina2018,Alkurdi2020}. Using a surface perturbation approach, Pendry \textit{et al.} \cite{Pendry2016} predicted that the heat transfer coefficient due to acoustic phonon exceeds that obtained with fluctuational electrodynamics for vacuum gaps smaller than 0.4 nm. Alkurdi \textit{et al.} \cite{Alkurdi2020} compared the contributions of radiation, phonons and electrons to heat transfer between Au surfaces separated by vacuum gaps varying from 1.5 nm down to 0.2 nm. Using a three-dimensional (3D) lattice dynamics framework, it was found that acoustic phonon transport exceeds radiation for all vacuum gaps considered, whereas electron tunneling slightly surpasses the phonon contribution for vacuum gaps smaller than 0.4 nm. In stark contrast, Messina {\it{et al.}} \cite{Messina2018} predicted that acoustic phonon transport does not play any role in extreme near-field heat transfer between Au surfaces. Their results suggested that for vacuum gaps smaller than $\sim$1 nm, electron tunneling largely dominates heat transport, whether or not a bias voltage is applied between the two surfaces.

Clearly, the mechanisms driving heat transfer between metallic surfaces separated by sub-10-nm vacuum gaps are not well understood. The objective of this work is to elucidate the physics underlying extreme near-field heat transfer between Au surfaces, and to determine vacuum gap ranges in which radiation, phonon, and electron are the dominant carriers. This is achieved by implementing a comprehensive model  that accounts for all energy carriers, and by comparing theoretical predictions against the experimental data from Refs. \cite{Kim2015,Kloppstech2017,Messina2018}. It is shown that in the absence of bias voltage, thermal transport is mediated by acoustic phonon transport for vacuum gaps smaller than $\sim$2 nm. In addition, the results demonstrate that the application of a bias voltage enhances the contributions of acoustic phonon and electron tunneling owing to the long-range Coulomb force and the lower potential barrier, respectively. 

To compare the theory against experimental results, the thermal conductance $G$ between a tip and a surface separated by a vacuum gap of thickness $d$ is calculated from local heat transfer coefficients $h$ between two parallel surfaces, modeled as semi-infinite layers, using the Derjaguin approximation \cite{Rousseau2009,Derjaguin1934a}: 
\begin{equation}
    G  \ = \ \int_{0}^{r_{\mathrm{tip}}}dr h(\tilde{d}){2\pi}r
    \label{Eq:derjaguin}
\end{equation} 
where $r$ is the radial direction, $r_{\mathrm{tip}}$ is the tip radius, and $\tilde{d} = d + r_{\mathrm{tip}} - \sqrt{{r_{\mathrm{tip}}}^2-r^2}$ is the vacuum gap distance between the surfaces (see Fig. 1). The heat transfer coefficient in Eq. (\ref{Eq:derjaguin}) is the sum of contributions from radiation, phonon, and electron, as detailed hereafter. 

\begin{figure}[p!]
\centering
\vspace{200pt}
\includegraphics[width=0.8\linewidth]{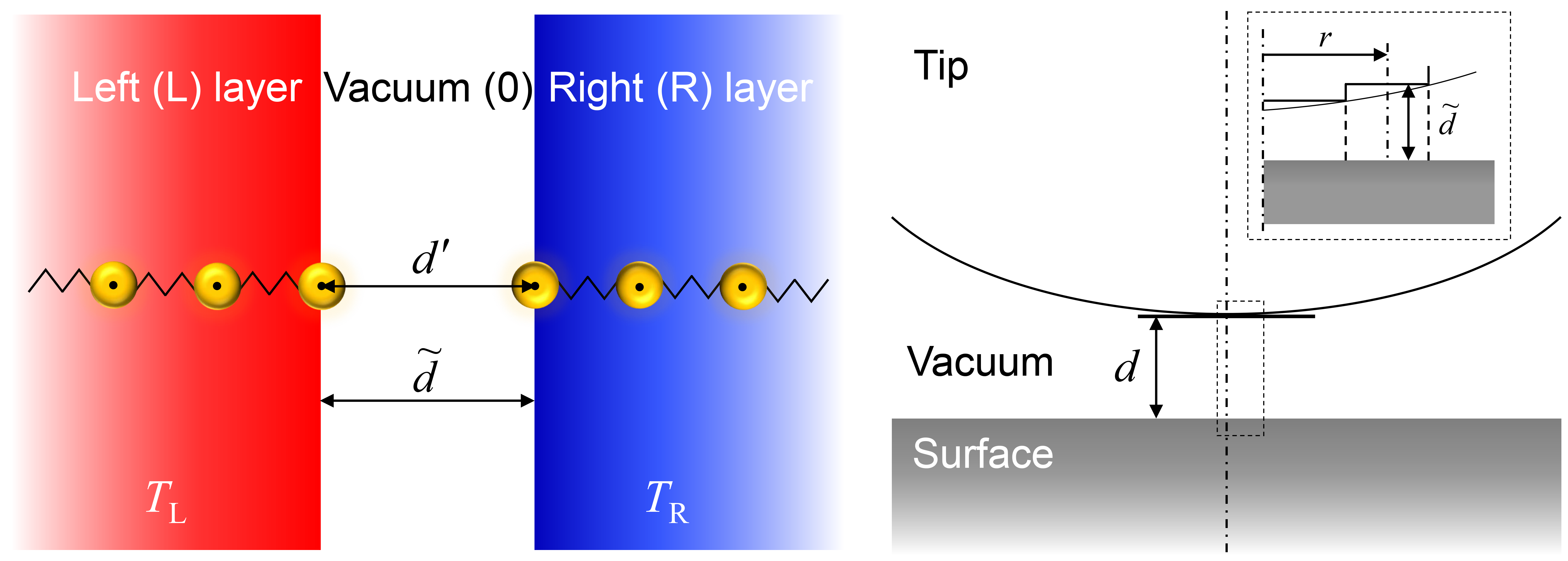}
\caption{\footnotesize{The thermal conductance between a tip and a surface separated by a vacuum gap $d$ is calculated from the heat transfer coefficients between two parallel surfaces, modeled as semi-infinite layers, separated by a vacuum gap $\tilde{d}$ using the Derjaguin approximation. Top-most atoms are located at the Au-vacuum interfaces. The interatomic vacuum distance $d^{\prime}$ used for predicting acoustic phonon transport is the same as the vacuum gap thickness $\tilde{d}$ used for radiation and electron tunneling calculations. The left (L) and right (R) semi-infinite layers are assumed to be at constant and uniform temperatures $T_\mathrm{L}$ and $T_\mathrm{R}$.}}
\label{Fig.1}
\end{figure}

The heat transfer coefficient due to radiation between two parallel surfaces (media L and R) separated by a vacuum gap of thickness $\tilde{d}$ (medium 0) is calculated using fluctuational electrodynamics to account for the near-field effects \cite{Polder1971, Rytov1987}: 
\begin{equation}
\label{radiative flux}
h_\mathrm{rad}  \ = \ \frac{1}{\pi^{\mathrm{2}}(T_\mathrm{L}-T_\mathrm{R})}\int_{0}^{\infty}d{\omega}\left[	\Theta\left(\omega,T_\mathrm{L}\right) - \Theta\left(\omega,T_\mathrm{R}\right)\right]
\\ \\ 
\int_{0}^{\infty}dk_\mathrm{\rho}k_\mathrm{\rho}\sum_{\gamma=\mathrm{TE,TM}}{\cal T_\mathrm{rad}^{\gamma}}\left(\omega,k_\mathrm{\rho}\right)
\end{equation} 
where $\omega$ is the angular frequency, $k_\mathrm{\rho}$ is the component of the wavevector parallel to an interface, and $\Theta(\omega, T)$ is the mean energy of an EM state calculated as  $\hbar\omega/[\mathrm{exp}({\hbar\omega/k_\mathrm{b}T})-1]$. The transmission functions in polarization state $\gamma$ for propagating ($k_\mathrm{\rho}$ $\leq$ $k_\mathrm{0}$) and evanescent ($k_\mathrm{\rho}$ $>$ $k_\mathrm{0}$) EM waves in vacuum are respectively given by:
\begin{equation}
\label{propagating transmission}
{\cal T_\mathrm{rad,prop}^\gamma}\left(\omega,k_\mathrm{\rho}\right) \ = \frac{\left(1-\left|r_\mathrm{0L}^{\gamma}\right|^{2}\right)\left(1-\left|r_\mathrm{0R}^{\gamma}\right|^{2}\right)}{4\left|1-r_\mathrm{0L}^{\gamma}r_\mathrm{0R}^{\gamma}e^{2i\mathrm{Re}\left(k_{\mathrm{z0}}\right)\tilde{d}}\right|^{2}}
\end{equation}
\begin{equation}
\label{evanescent transmission}
{\cal T_\mathrm{rad,evan}^\gamma}\left(\omega,k_\mathrm{\rho}\right) \ = e^{-2\mathrm{Im}\left(k_\mathrm{z0}\right)\tilde{d}}\frac{\mathrm{Im}\left(r_\mathrm{0L}^{\gamma}\right)\mathrm{Im\left(r_\mathrm{0R}^{\gamma}\right)}}{\left|1-r_\mathrm{0L}^{\gamma}r_\mathrm{0R}^{\gamma}e^{-2\mathrm{Im}\left(k_{\mathrm{z0}}\right)\tilde{d}}\right|^{2}}
\end{equation} where $k_\mathrm{z0}$ is the component of the vacuum wavevector perpendicular to an interface, and $r_{0j}^\gamma$ ($j = \mathrm{L}, \mathrm{R}$) is the Fresnel reflection coefficient \cite{Yeh1988}. The frequency-dependent dielectric function of Au is calculated using the Drude model provided in Ref. \cite{Chapuis2008}.

Acoustic phonons can tunnel across vacuum gaps via force interactions. Acoustic phonon transport is modeled using the one-dimensional (1D) atomistic Green's function (AGF) method \cite{Zhang2007, Sadasivam2014}. The heat transfer coefficient due to acoustic phonon transport across an interatomic vacuum distance $d^\prime$ for a 1D atomic chain is written as:
\begin{equation}
\label{phonon heat flux}
h_\mathrm{ph}  \ = \ \frac{1}{A(T_\mathrm{L}-T_\mathrm{R})}\int_{0}^{\infty}d{\omega}\frac{\hbar\omega}{2\pi}{\cal T_\mathrm{ph}}({\omega})[N(\omega, T_\mathrm{L})-N(\omega,T_\mathrm{R})]
\end{equation} where $N = 1/[\mathrm{exp}({\hbar\omega/k_\mathrm{b}T})-1]$ is the Bose-Einstein distribution function and $A$ is the cross-sectional area of an atom. The atomic radius of Au is taken as 1.740 {\AA} \cite{Clementi1967}. The top-most atoms are located at the Au-vacuum interfaces \cite{Nakamura1997}, such that the interatomic vacuum distance $d^\prime$ is the same as the vacuum gap thickness $\tilde{d}$ used for radiation and electron tunneling calculations (see Fig. 1). Note that the lattice constant of Au (4.065 {\AA}) is used as the criterion for bulk contact. The minimum distance at which the heat transfer coefficient is calculated is therefore 4.065 {\AA}. The phonon transmission function is given by the Caroli formula \cite{Caroli1971}:
\begin{equation}
\label{transmission function}
{\cal T_\mathrm{ph}}({\omega}) \ = \ \mathrm{Trace}[\Gamma_\mathrm{L}G_\mathrm{d}\Gamma_\mathrm{R}G_\mathrm{d}^\mathrm{\dagger}]
\end{equation} 
where the superscript $\dagger$ denotes conjugate transpose, $\Gamma_\mathrm{L,R}$ is the escape rate of phonons from the device region to the semi-infinite layers, and $G_\mathrm{d}$ is the Green's function of the device region. In the present study, the device region encompasses the vacuum gap and five atoms in each of the semi-infinite layers. Increasing the number of atoms in the device region does not affect the phonon transmission function. Details about the 1D AGF method have been provided in Refs. \cite{Zhang2007,Sadasivam2014,Jarzembski2019}. 

For the case of two Au surfaces, the short- and long-range interactions across the vacuum gap are respectively described by the Lennard-Jones and Coulomb force models \cite{Chiloyan2015,Alkurdi2020}. The Lennard-Jones model accounts for the van der Waals force and overlapping electron cloud repulsive force, whereas the Coulomb force is effective only when there is a bias voltage \cite{Horn1992,Heinze1999,Properties1969}. The rationale behind the selection of these forces and the force models are provided in Sec. S1 of the Supplemental Material \cite{supplemental}. 

Electrons can contribute to heat transfer across vacuum gaps via tunneling and thermionic emission. Owing to the low temperatures considered ($\sim$300 K), the contribution of thermionic emission is negligible. The heat transfer coefficient due to electron tunneling across a vacuum gap can be written as \cite{Hishinuma2001,Westover2008,Wang2016}: 
\begin{equation}
\begin{split}
    h_\mathrm{el}  \ = \ \frac{1}{(T_\mathrm{L}-T_\mathrm{R})} \int_{-\infty}^{W_{\mathrm{max}}} dE_z[&\left(E_z + {k_\mathrm{b}}{T_\mathrm{L}}\right)N_\mathrm{L}\left(E_z,T_\mathrm{L}\right) 
    \\ 
    - &\left(E_z + k_\mathrm{b}{T_\mathrm{R}}\right)N_\mathrm{R}\left(E_z-eV_\mathrm{bias},T_\mathrm{R}\right)]{\cal T_{\mathrm{el}}}\left(E_z\right)
    \label{Eq:qel}
\end{split}
\end{equation} where $e = 1.602{\times}10^{-19} {\,} \mathrm{C}$ is the electron charge, $V_\mathrm{bias}$ is the bias voltage, $E_z$ is the electron energy perpendicular to the surfaces, and $W_\mathrm{max}$ is the maximum potential barrier. The term $N_j(E_z, T_j)$, denoting the number of electrons at energy level $E_z$ per unit area and per unit time, is calculated as \cite{Hishinuma2001}: 
\begin{equation}
    N_j\left(E_{z}, T_j\right)  \ = \ \frac{m_\mathrm{e}k_\mathrm{b}T_j}{2{\pi}^2\hbar^3}\mathrm{ln}\left[1 + \mathrm{exp}\left(-\frac{E_z-E_{F,j}}{k_\mathrm{b}T_j}\right)\right]
    \label{Eq:numberofelectron}
\end{equation} where $m_\mathrm{e} = \mathrm{9.109\times10^{-31}{\,}kg}$ is the electron mass, and $E_{F,j}$ is the Fermi level used as a reference for computing the electron energy \cite{Hishinuma2001}. The electron transmission function is calculated using the Wentzel-Kramers-Brillouin (WKB) approximation \cite{Berry1972}:
\begin{equation}
    {\cal T_{\mathrm{el}}}\left(E_z\right)  \ = \ \mathrm{exp}\left[-\frac{\sqrt{8m_\mathrm{e}}}{\hbar}\int_{z_1}^{z_2}dz \sqrt{W\left(z\right)-E_z} \right]
\end{equation}
where $W(z)$ is the potential barrier profile in the vacuum gap, while $z_1$ and $z_2$ are the roots of $W(z)-E_z=0$ delimiting the width of the electron tunneling barrier $E_z$. The potential barrier profile can be expressed as \cite{Hishinuma2001}:
\begin{equation}\label{profile_hishinuma}
    W(z) = W_{id}(z) + W_{ic}(z)  
\end{equation} Note that the space-charge effect is assumed to be fully suppressed for sub-10-nm vacuum gaps \cite{Wang2016}. The ideal barrier profile \cite{Wang2016} and image-charge perturbation \cite{Baldea2012} are respectively calculated as:
\begin{equation}
    W_{id}(z) = \Phi_\mathrm{L} - (\Phi_\mathrm{L} - \Phi_\mathrm{R} - eV_\mathrm{bias})\left(\frac{z}{\tilde{d}}\right)
\end{equation}

\begin{equation}
    W_{ic}(z) = \frac{e^{2}}{16{\pi}{\epsilon}_{0}d}\left[-2{\Psi}\left(1\right)+{\Psi}\left(\frac{z}{\tilde{d}}\right)+{\Psi}\left(1-\frac{z}{\tilde{d}}\right)\right]
\end{equation} where $\Phi_{\mathrm{L}, \mathrm{R}} = 5.10{~}\mathrm{eV}$ is the work function of Au \cite{Zhou2012} and $\Psi$ is the digamma function.

The WKB approximation may lead to an overestimation of the electron heat transfer coefficient below vacuum gaps of $\sim$0.5 nm \cite{Olesen1996,Westover2008}. As such, for small vacuum gaps, the electron heat transfer coefficient is also calculated via the 1D non-equilibrium Green's function (NEGF) approach described in Ref. \cite{Westover2008}. In the NEGF framework, the electron transmission function is given by:
\begin{equation}
\label{transmission function NEGF}
{\cal T_\mathrm{el}}(E_{z}) \ = \ \mathrm{Trace}[\Gamma_\mathrm{L}G^\mathrm{R}\Gamma_\mathrm{R}\left(G^\mathrm{R}\right)^\mathrm{\dagger}],
\end{equation} where $\Gamma_\mathrm{L,R}$ is the energy broadening matrix, whereas $G^\mathrm{R}$ is the retarded Green's function. 

\begin{figure}[p!]
\centering
\vspace{150pt}
\includegraphics[width=1\linewidth]{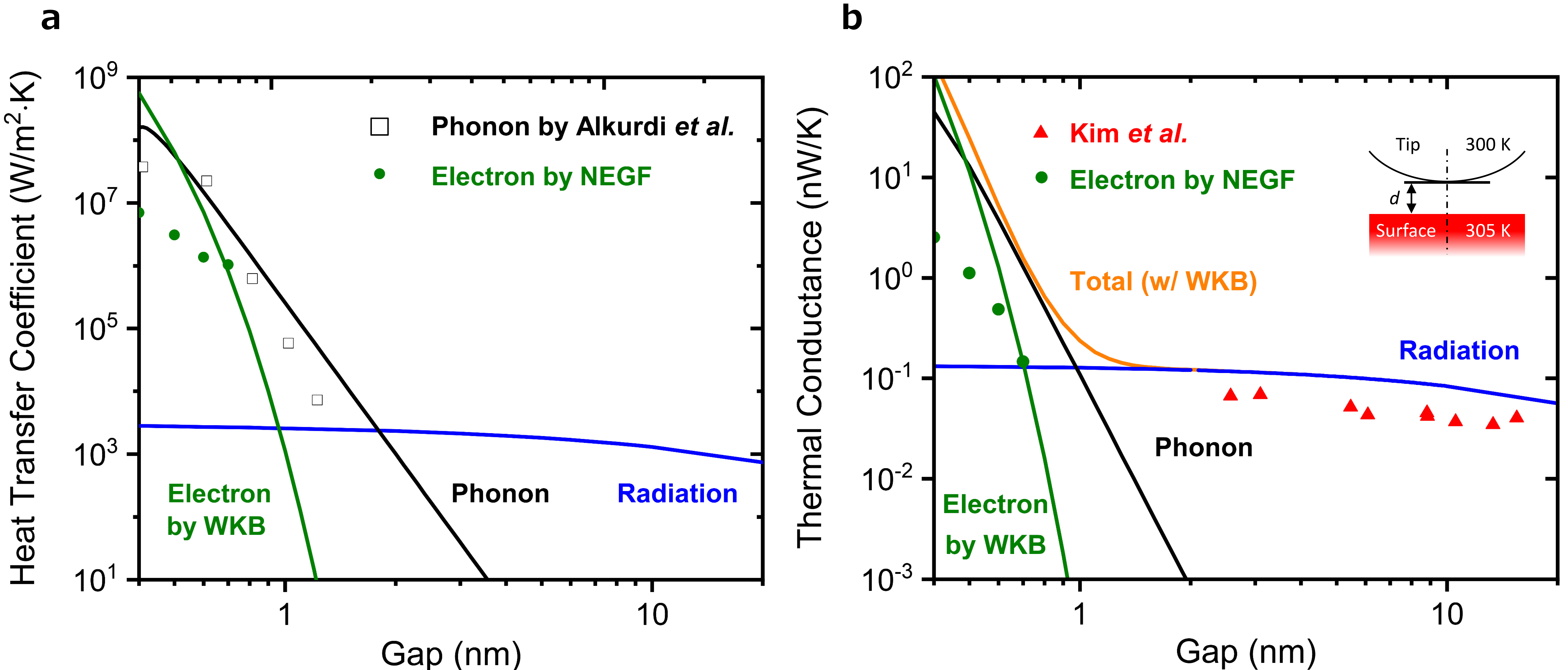}
\caption{\footnotesize{(a) Heat transfer coefficients due to radiation, phonon, and electron between two Au surfaces separated by a vacuum gap $\tilde{d}$ ($T_\mathrm{{L}}$ = 305 K, $T_\mathrm{{R}}$ = 300 K, $V_\mathrm{bias}$ = 0 V). The electron heat transfer coefficient is calculated via the WKB approximation and the NEGF. The phonon heat transfer coefficient predicted via the 1D AGF method is compared against the 3D lattice dynamics results of Alkurdi \textit{et al.} \cite{Alkurdi2020}. (b) Thermal conductance between a 450-nm-radius tip (300 K) and a surface (305 K), both made of Au, separated by a vacuum gap $d$. The total conductance is calculated using the electron heat transfer coefficient from the WKB approximation. The predicted conductance is compared against the experimental results of Kim \textit{et al.} \cite{Kim2015}.}}
\label{Fig.2}
\end{figure}

The heat transfer coefficients between two Au surfaces due to radiation, acoustic phonon transport, and electron tunneling are presented in Fig. 2(a) for the experimental conditions taken from Kim \textit{et al.} \cite{Kim2015}, where $T_\mathrm{{L}}$ = 305 K, $T_\mathrm{{R}}$ = 300 K, and $V_\mathrm{bias}$ = 0 V. The Au-vacuum interfaces support surface plasmon polaritons at a high frequency ($\sim\mathrm{1.2\times10^{16}}$ rad/s) that cannot be thermally excited at room temperature. As such, the radiation heat transfer coefficient saturates as the vacuum gap decreases \cite{Chapuis2008}. It should be noted that while fluctuational electrodynamics is not expected to be valid for sub-2-nm vacuum gaps owing to non-local effects \cite{Chapuis2008}, radiation predictions below 2 nm vacuum gaps are shown throughout the paper for reference. The accuracy of the 1D AGF approach for predicting acoustic phonon transport between Au surfaces is verified in Fig. 2(a) by comparison against the results obtained from a 3D lattice dynamics framework \cite{Alkurdi2020}. In the absence of bias voltage, only the van der Waals and overlapping electron cloud forces, represented by the Lennard-Jones model, contribute to acoustic phonon transport. The phonon and electron heat transfer coefficients exceed that of radiation for vacuum gaps smaller than $\sim$2 nm and $\sim$1 nm, respectively. Provided that the NEGF calculates the electron transmission function more accurately than the WKB approximation for vacuum gaps smaller than $\sim$0.5 nm, the results indicate that the phonon contribution always exceed that of the electron for sub-2-nm vacuum gaps near room temperature.

Figure 2(b) shows the predicted thermal conductance between a 450-nm-radius Au tip at 300 K and an Au surface at 305 K by applying the Derjaguin approximation with the heat transfer coefficients shown in Fig. 2(a). These predictions are compared against the conductance measured in Ref. \cite{Kim2015} for vacuum gaps down to $\sim$2 nm. Clearly, for vacuum gaps larger than $\sim$1.5 nm, heat transfer is mediated by radiation. The calculated total conductance is in good agreement with the experimental data of Ref. \cite{Kim2015}. The results from the comprehensive model presented in this work therefore support the conclusions of Kim \textit{et al.} \cite{Kim2015}. Vacuum gaps approximately equal to or smaller than 1 nm would have been required to observe a significant enhancement of the conductance due to acoustic phonon transport.  

Figure 3(a) presents the heat transfer coefficients between two Au surfaces due to radiation, acoustic phonon transport, and electron tunneling for the experimental conditions taken from Kloppstech {\it{et al}}. \cite{Kloppstech2017}, where $T_\mathrm{L}$ = 280 K, $T_\mathrm{R}$ = 120 K, and $V_\mathrm{bias}$ = 0.6 V. It should be noted that an electric field generated by a bias voltage not only enhances electron tunneling but also induces surface charges \cite{Properties1969,Heinze1999}, which are at the origin of the Coulomb force \cite{Horn1992}. Since the long-range Coulomb force can drive acoustic phonon transport across vacuum gaps \cite{Chiloyan2015}, the surface charge density under a bias voltage should be determined and taken into account for the calculations. Jarzembski {\it{et al}}. \cite{Jarzembski2019} estimated the surface charge density of a biased platinum surface based on their experimental conditions. A surface charge density of $\mathrm{8\times10^{-4} ~C/{m^2}}$ was estimated for a bias voltage of 0.8 V. In the present work, a value of $\mathrm{6\times10^{-4} ~C/{m^2}}$ is used under the assumption that the surface charge density varies linearly with the bias voltage \cite{Chen1999, Liu2020}. Note that the phonon heat transfer coefficient is calculated with and without bias voltage, and the shaded area in Fig. 3(a) shows its possible values.

\begin{figure}[p!]
\centering
\vspace{150pt}
\includegraphics[width=1\linewidth]{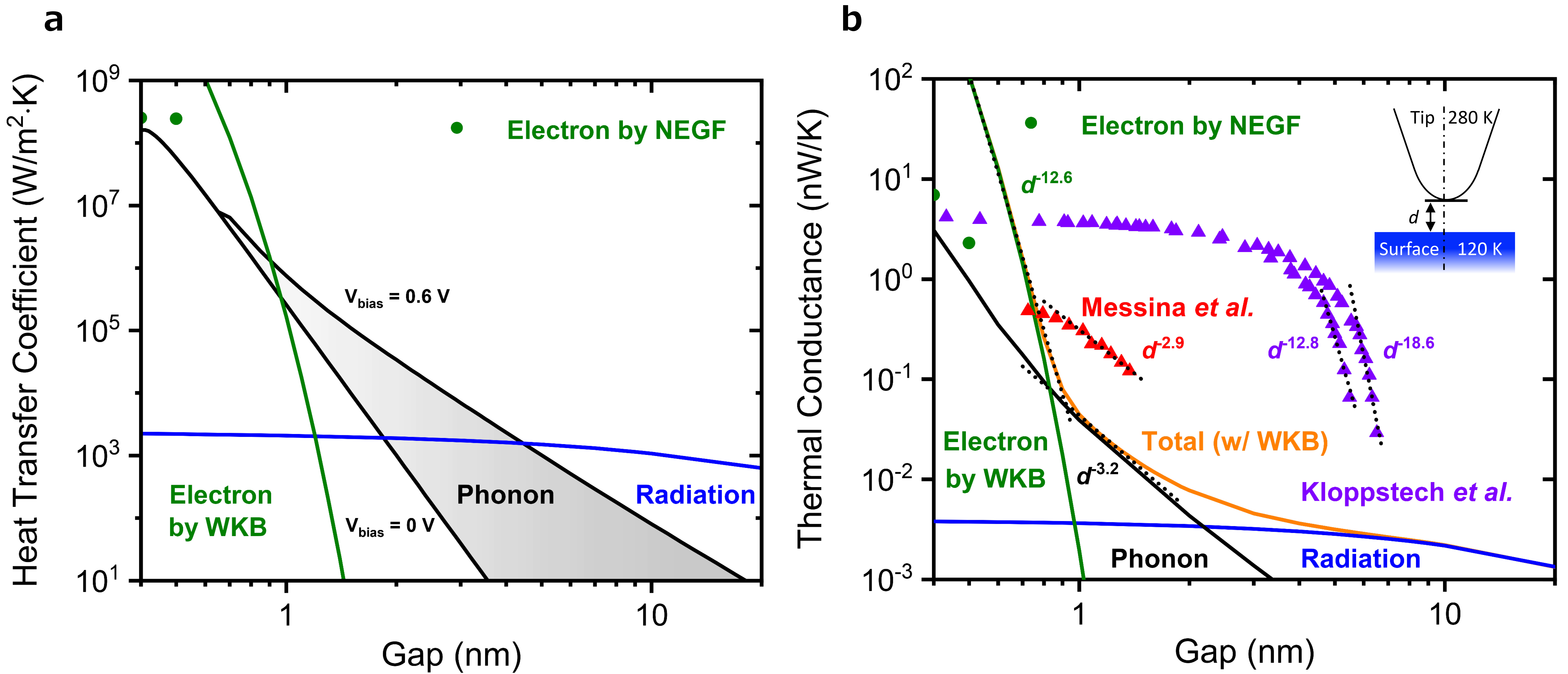}
\caption{\footnotesize{(a) Heat transfer coefficients due to radiation, phonon, and electron between two Au surfaces separated by a vacuum gap $\tilde{d}$ ($T_\mathrm{{L}}$ = 280 K, $T_\mathrm{{R}}$ = 120 K, $V_\mathrm{bias}$ = 0.6 V). The electron heat transfer coefficient is calculated via the WKB approximation and the NEGF. The phonon heat transfer coefficient is calculated both with and without bias voltage. (b) Thermal conductance between a 30-nm-radius tip (280 K) and a surface (120 K), both made of Au, separated by a vacuum gap $d$. The total conductance is calculated using the electron heat transfer coefficient from the WKB approximation. The predicted conductance is compared against the experimental results of Kloppstech \textit{et al.} \cite{Kloppstech2017} and Messina \textit{et al.} \cite{Messina2018}.}}
\label{Fig.3}
\end{figure}

Owing to the long-range Coulomb force, the phonon heat transfer coefficient exceeds that of radiation for vacuum gaps smaller than $\sim$5 nm. The Coulomb force, however, has a low impact on the phonon heat transfer coefficient for vacuum gaps smaller than 1 nm (see Sec. S1 of the Supplemental Material for more details \cite{supplemental}). The electron heat transfer coefficient is enhanced due to the bias voltage that lowers the potential barrier for tunneling. The WKB approximation suggests that electron-mediated heat transfer exceeds acoustic phonon transport for vacuum gaps smaller than 1 nm. The NEGF also suggests that the electron heat transfer coefficient is slightly larger than that of phonon, but for sub-0.5-nm vacuum gaps. 

The thermal conductance calculated between a 30-nm-radius Au tip and an Au surface is shown in Fig. 3(b) under the same experimental conditions as in Fig. 3(a) (temperatures of 280 and 120 K, $V_\mathrm{bias}$ = 0.6 V). The results are compared against the experimental data of Kloppstech {\it{et al.}} \cite{Kloppstech2017}, in addition to a more recent set of data from the same group (Messina {\it{et al.}} \cite{Messina2018}). Note that the most recent experimental data were obtained for surface and tip temperatures of respectively 195 K and 295 K. These slight temperature differences have no noticeable impact on the heat transfer coefficients (see Sec. S2 of the Supplemental Material \cite{supplemental}). The total conductance, which include the radiation, phonon, and electron contributions, display three distinct regimes. For vacuum gaps larger than 5 nm, radiation transport via evanescent EM waves drives heat transfer. For vacuum gaps from 5 nm down to $\sim$1.5 nm, radiation transitions to acoustic phonon-mediated heat transfer. Acoustic phonon transport becomes dominant between vacuum gaps of 1.5 nm and 0.9 nm. For vacuum gaps smaller than 0.9 nm, electron tunneling drives heat transfer owing to the applied bias voltage. The magnitude of the experimental and theoretical data are quite different. Nevertheless, a power law analysis may enable determining the energy carriers driving heat transport in the experiments of Refs. \cite{Kloppstech2017,Messina2018}. In the vacuum gap range of 1 nm to 1.5 nm where acoustic phonon transport is dominant, the predicted total conductance follows a $d^{-3.2}$ power law. Within that range, the data of Messina {\it{et al.}} \cite{Messina2018} vary as $d^{-2.9}$, which is close to the theory. This regime arises due to the Coulomb force induced by the bias voltage. For vacuum gaps smaller than 0.9 nm where electron tunneling dominates heat transport, the total conductance follows a $d^{-12.6}$ power law. Interestingly, a similar trend is observed in the experimental data of Kloppstech {\it{et al.}} \cite{Kloppstech2017} but at much larger vacuum gaps ($d\gtrsim$ 5 nm).

\begin{figure}[p!]
\centering
\vspace{150pt}
\includegraphics[width=0.7\linewidth]{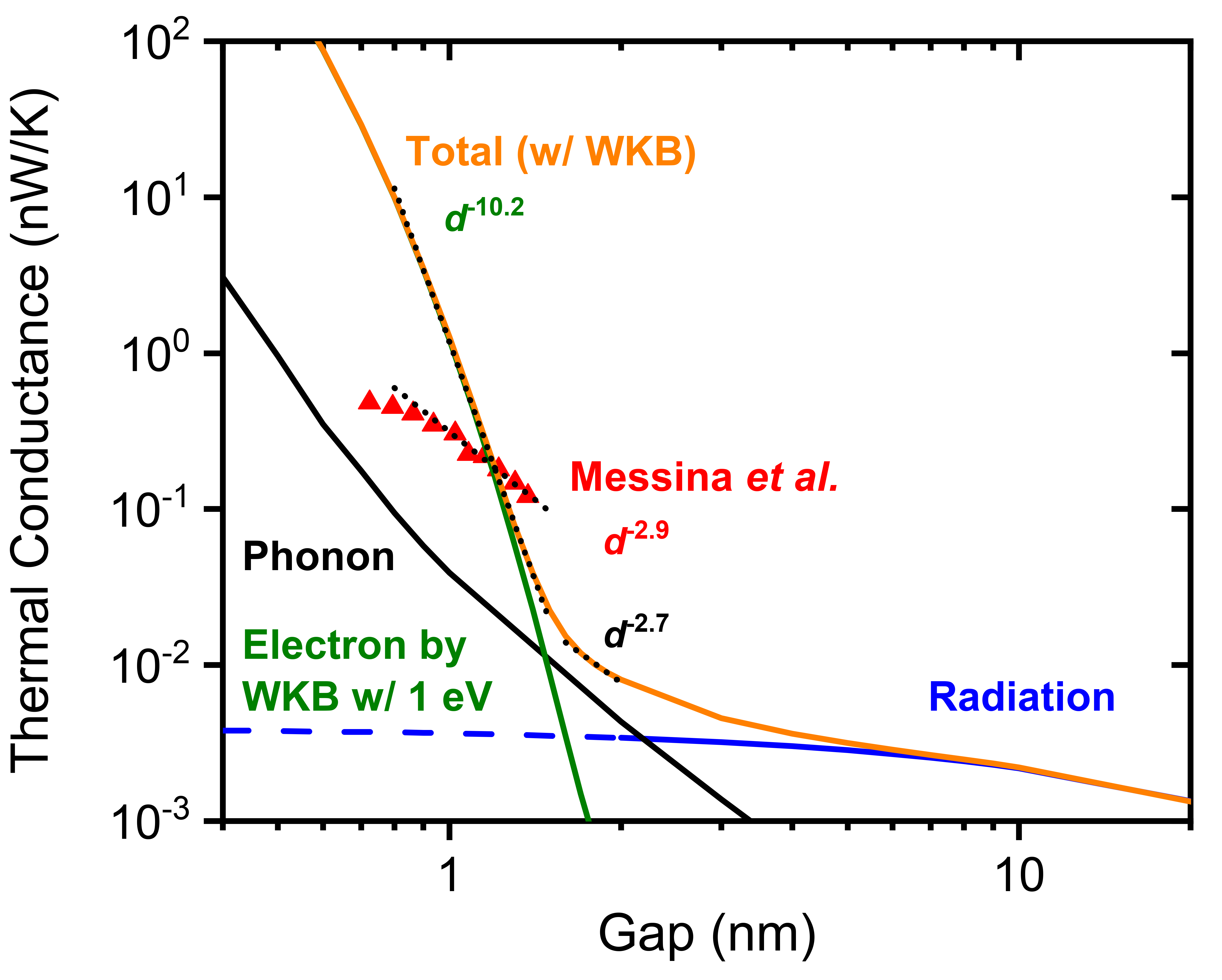}
\caption{\footnotesize{Thermal conductance between a 30-nm-radius tip (280 K) and a surface (120 K), both made of Au, separated by a vacuum gap $d$. The electron heat transfer coefficient is calculated with the WKB approximation using an apparent potential barrier of 1 eV. The phonon heat transfer coefficient is calculated with a bias voltage of 0.6 V. The predicted conductance is compared against the experimental results of Messina \textit{et al.} \cite{Messina2018}.}}
\label{Fig.4}
\end{figure}

It has been hypothesized that contaminants bridging the tip and the surface prior to bulk contact may lower the potential barrier for electron tunneling \cite{Cui2017}. Using their measured tunneling current, Messina {\it{et al.}} \cite{Messina2018} estimated an apparent potential barrier of 1 eV, which induces an enhancement of the electron heat transfer coefficient. Fig. 4 shows the thermal conductance predicted by considering a simple potential barrier of 1 eV for electron tunneling. As expected, electron tunneling for  that case becomes dominant at a larger vacuum gap ($\sim$1.5 nm). The magnitude of the total conductance is closer to the experimental data of Messina {\it{et al.}} \cite{Messina2018}, but the trends are different. Electron-mediated heat transport follows a $d^{-10.2}$ power law, which is not experimentally observed. In the gap range between 1.5 nm and 2 nm where all energy carriers contribute to the conductance, a $d^{-2.7}$ power law is predicted. This is again close to the $d^{-2.9}$ power law of Messina {\it{et al.}} \cite{Messina2018} which however arises in a slightly different vacuum gap range (1 nm to 1.5 nm).
It is concluded that the measured conductance reported by Messina {\it{et al.}} \cite{Messina2018} is not solely due to electron tunneling. Heat transport for that case is likely to be a combination of acoustic phonon transport, enhanced via the bias voltage, and electron tunneling possibly enhanced by the low apparent potential barrier mediated by contaminants. It has been pointed out that the data of Kloppstech {\it{et al.}} \cite{Kloppstech2017} were obtained at different apparent potential barrier heights \cite{Messina2018}, which make their interpretation challenging. Yet, the variation of the conductance with respect to the vacuum gap suggests electron-mediated heat transfer.

In summary, a comprehensive model combining fluctuational electrodynamics for radiation, the AGF method for phonon, and the WKB approximation in addition to the NEGF approach for electron has been implemented to analyze extreme near-field heat transfer between Au surfaces. In the absence of bias voltage and near room temperature, heat transfer is mediated by radiation through evanescent EM waves down to a vacuum gap of $\sim$2 nm, below which acoustic phonon transport is the dominant energy carrier. It was shown that the application of a bias voltage increases not only electron tunneling, but also acoustic phonon transport mediated by the long-range Coulomb force. As a result, for a bias  voltage of 0.6 V, acoustic phonon transport can affect heat transfer for vacuum gaps from 5 nm down to $\sim$1 nm, and electron tunneling should be a responsible heat transfer mechanism below 1 nm. However, comparison of the comprehensive model against state-of-the-art measurements \cite{Kim2015,Kloppstech2017,Messina2018} reveals that additional data are needed to experimentally quantify the contributions of acoustic phonon and electron as a function of the bias voltage in the heat transfer between Au surfaces separated by sub-10-nm vacuum gaps. 

This work was supported by the National Science Foundation (grants CBET-1605584 and CBET-1952210). T.T. also appreciate the financial support by the Yamada Science Foundation and the Fujikura Foundation.

\normalem 
\bibliography{Ref}
\end{document}



\title{Supplemental Material \\Extreme Near-Field Heat Transfer Between Gold Surfaces}

\author{Takuro Tokunaga}
\affiliation{
Department of Mechanical Engineering, University of Utah, Salt Lake City, Utah 84112, United States.}
 
\author{Amun Jarzembski}
\affiliation{
Sandia National Laboratories, Albuquerque, NM 87185, United States.}

\author{Takuma Shiga}
\affiliation{
Department of Mechanical Engineering, The University of Tokyo, Bunkyo, Tokyo 113-8656, Japan.}

\author{Keunhan Park}
\email{kpark@mech.utah.edu}
\affiliation{
Department of Mechanical Engineering, University of Utah, Salt Lake City, Utah 84112, United States.}

\author{Mathieu Francoeur}
\email{mfrancoeur@mech.utah.edu}
\affiliation{
Department of Mechanical Engineering, University of Utah, Salt Lake City, Utah 84112, United States.}



\maketitle

\section{S1. LENNARD-JONES AND COULOMB FORCE MODELS}
Short- and long-range force interactions are considered for modeling acoustic phonon transport across vacuum gaps. The short-range interactions include the repulsive force due to overlapping electron clouds and the van der Waals force \cite{Sadewasser2003}. The possible long-range interactions are the Maxwell stress of the electric field (capacitance force) \cite{Pendry2016}, the Maxwell stress of the magnetic field \cite{Pendry2016}, and the Coulomb force \cite{Chiloyan2015}. 

The short-range force interactions are modeled via the first order derivative of the Lennard-Jones potential, $U_\mathrm{L-J}$, with respect to the interatomic distance, $d^{'}$ \cite{Jones1924}: 
\begin{equation}
\label{L-J force}
F_\mathrm{L-J}  \ = \ \frac{\partial U_\mathrm{L-J}}{\partial d^{'}} = -24\left(\frac{\varepsilon}{d^{'}}\right)\left[2\left(\frac{\sigma}{d^{'}}\right)^{12} - \left(\frac{\sigma}{d^{'}}\right)^{6}\right]
\end{equation}
where $\varepsilon = 8.55{\times}10^{-20}{\ }\mathrm{J}$ and $\sigma = 2.57{\times}10^{-10}{\ }\mathrm{m}$ \cite{Yu2004}. In Eq. (\ref{L-J force}), the term varying as $d^{'-12}$ describes the overlapping electron cloud repulsive force \cite{Senior1924, Mohebifar2017}, whereas the $d^{'-6}$ term describes the van der Waals force interactions \cite{Senior1924, Li2003}.

For the long-range force interactions, the Maxwell stress of the magnetic field does not affect low-frequency phonon heat transfer and can thus be ignored \cite{Pendry2016}. The application of a bias voltage induces surface charges that in turn generate capacitance and Coulomb forces \cite{Horn1992, Heinze1999, Lang1973}. The capacitance force induced by a bias voltage between a tip and a surface is given by \cite{Terris1989}:
\begin{equation}
\label{capacitance}
 \overline{F}_\mathrm{Capacitance}  \ = \ \frac{\epsilon_0V_{\mathrm{bias}}}{2}\frac{A_\mathrm{{tip}}}{d}
\end{equation} where $\epsilon_0$ is the permittivity of free space, $V_\mathrm{bias}$ is the bias voltage, $A_\mathrm{tip}$ is the tip surface, and $d$ is the vacuum gap thickness. Eq. (\ref{capacitance}) describes the capacitance force over the entire tip surface. For acoustic phonon transport calculations based on the atomistic Green's function method, the force constants between gold (Au) atoms must be calculated. The capacitance force between Au atoms is obtained by normalizing Eq. (\ref{capacitance}) by the tip surface, $A_\mathrm{tip}$, and by then multiplying the resulting expression by the cross-sectional area of an Au atom, $A$: 
\begin{equation}
\label{capacitance atom}
 F_\mathrm{Capacitance}  \ = \ \frac{\epsilon_0V_{\mathrm{bias}}}{2}\frac{A}{d^{'}}
\end{equation} where the vacuum gap $d$ has been replaced by the interatomic distance $d^{'}$.

The Coulomb force due to surface charges between a tip and a surface is calculated as follows \cite{Terris1989}:
\begin{equation}
\label{coulomb}
 \overline{F}_\mathrm{Coulomb}  \ = \ \frac{Q_\mathrm{s}Q_\mathrm{t}}{4{\pi}{\epsilon_0}{d^{2}}}
\end{equation} where $Q_\mathrm{s}$ is the surface charge, and $Q_\mathrm{t}$ is the tip charge defined as $Q_\mathrm{t} = -(Q_\mathrm{s} + Q_\mathrm{e})$. The induced capacitive charge is given by $Q_\mathrm{e} = CV_{\mathrm{bias}}$, where $C$ ($= \epsilon_{0}{A_{\mathrm{tip}}}/d^{2}$) is the parallel plate capacitance \cite{ElKhoury2016}. Following the same procedure as for the capacitance force, the Coulomb force between Au atoms is given by:  
\begin{equation}
\label{coulomb atom}
 F_\mathrm{Coulomb}  \ = \ \frac{Q_\mathrm{s}Q_\mathrm{t}}{4{\pi}{\epsilon_0}{d^{'2}}}\frac{A}{{A_\mathrm{tip}}}
\end{equation} where the vacuum gap $d$ has been replaced by the interatomic distance $d^{'}$. 

\begin{figure}[H]
\centering
\includegraphics[width=0.7\linewidth]{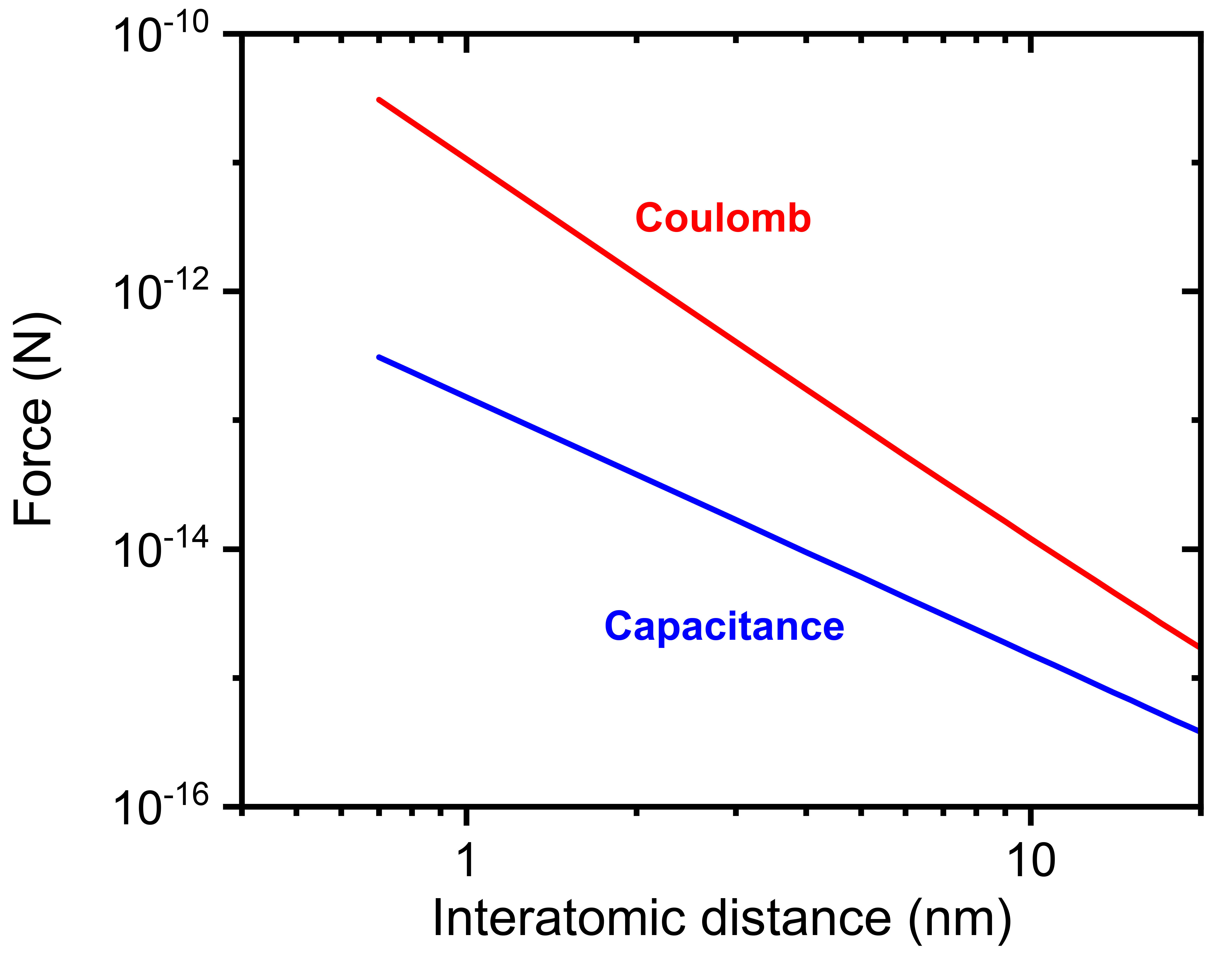}
\caption{\footnotesize{Comparison of the capacitance and Coulomb forces between Au atoms ($V_{\mathrm{bias}}$ = 0.6 V, $Q_{\mathrm{s}}/A_{\mathrm{tip}}$ = $\mathrm{6\times10^{-4}}$ $\mathrm{C/m^2}$, 30-nm-radius tip).}}
\label{Figure: force}
\end{figure}

The capacitance and Coulomb forces between Au atoms, calculated via Eqs. (\ref{capacitance atom}) and (\ref{coulomb atom}), as a function of the interatomic vacuum distance, $d^{'}$, are compared in Fig. S\ref{Figure: force} for a bias voltage, $V_\mathrm{bias}$, of 0.6 V and a surface charge density, $Q_{\mathrm{s}}/A_{\mathrm{tip}}$, of $\mathrm{6\times10^{-4}}$ $\mathrm{C/m^2}$. Note that the tip radius of 30 nm reported in Refs. \cite{Kloppstech2017,Messina2018} is considered in the simulations. 

 The Coulomb force exceeds the capacitance force for all interatomic vacuum distances. As such, the Coulomb force is the only long-range interactions considered for acoustic phonon transport calculations.

The total force driving acoustic phonon transport across vacuum gaps is written as:
\begin{equation}\label{total force}
F_\mathrm{total}  \ = \ F_\mathrm{L-J} + F_\mathrm{Coulomb}
\end{equation} The inputs for atomistic Green's function calculations are the force constants which are obtained by taking the absolute value of the first order derivative of Eq. (\ref{total force}) with respect to $d^{'}$ \cite{Ezzahri2014}.

At the onset of contact, the Coulomb force vanishes due to charge neutralization \cite{Srisonphan2014}. The reduction of the surface charge, $Q_\mathrm{s}$, with respect to the vacuum gap between a tip and a surface has been previously predicted in Refs. \cite{Hong1998,Behrens2001}. In addition, an experimental effort has demonstrated a vanishing capacitance between a tip and a substrate with the reduction of the vacuum gap distance \cite{Yoo1997}. As such, the surface charge should ideally be treated as a function of the interatomic distance \cite{Behrens2001}. However, since it is challenging to develop a gap-dependent surface charge model, $Q_\mathrm{s}$ is treated here as a constant value that vanishes below a cut-off vacuum distance \cite{Butt1991}. The cut-off value is approximated as the interatomic vacuum distance for which electron tunneling becomes significant, corresponding to ${d^{'}}{\,}\mathrm{\approx 7{\,}{\mbox{\AA}}}$ \cite{Anselmetti1994,McCarty2008}. 

\section{S2. IMPACT OF TEMPERATURE ON THE HEAT TRANSFER COEFFICIENT}

\begin{figure}[H]
\centering
\includegraphics[width=0.7\linewidth]{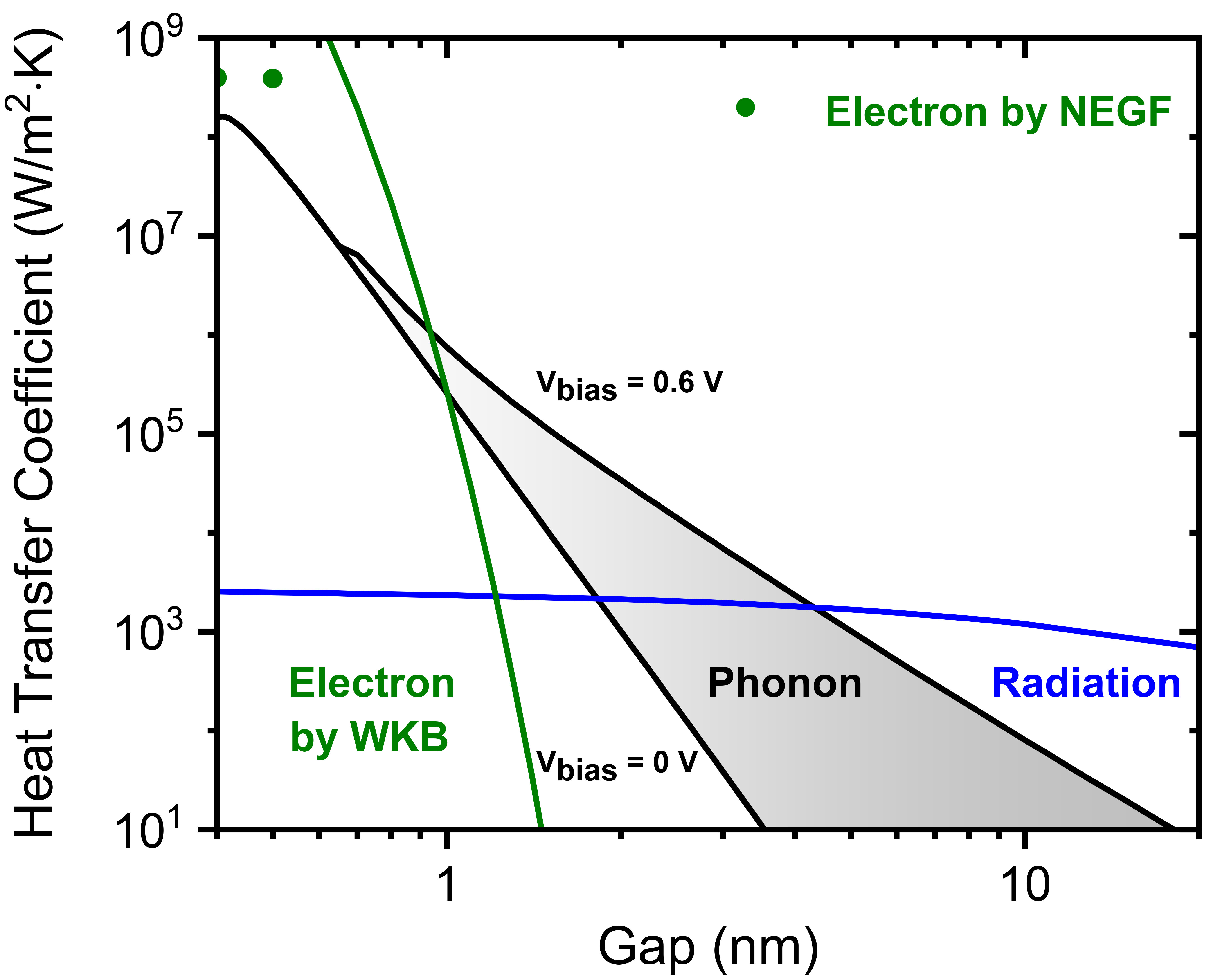}
\caption{\footnotesize{Heat transfer coefficients due to radiation, phonon, and electron between two Au surfaces separated by a vacuum gap $\tilde{d}$ ($T_{\mathrm{L}}$ = 295 K, $T_{\mathrm{R}}$ = 195 K , $V_{\mathrm{bias}}$ = 0.6 V). The electron heat transfer coefficient is calculated via the Wentzel-Kramers-Brillouin (WKB) approximation and the non-equilibrium Green's
function (NEGF) approach.}}
\label{HTC_different_temperature}
\end{figure}

The thermal conductance between a 30-nm-radius tip and a surface, calculated via the heat transfer coefficients of Fig. 3(a) ($T_\mathrm{L}$ = 280 K, $T_\mathrm{H}$ = 120 K, $V_\mathrm{bias}$ = 0.6 V) and the Derjaguin approximation, is reported in Fig. 3(b). The predicted conductance is compared against the experimental data of Kloppstech {\it{et al.}} ($T_\mathrm{L}$ = 280 K, $T_\mathrm{H}$ = 120 K, $V_\mathrm{bias}$ = 0.6 V) \cite{Kloppstech2017} in addition to those of Messina {\it{et al.}} ($T_\mathrm{L}$ = 295 K, $T_\mathrm{H}$ = 195 K, $V_\mathrm{bias}$ = 0.6 V) \cite{Messina2018}. Fig. S\ref{HTC_different_temperature} shows the heat transfer coefficients between two Au surfaces due to radiation, acoustic phonon transport, and electron tunneling for the experimental conditions of Messina {\it{et al.}} \cite{Messina2018}. It is clear by comparing Fig. S\ref{HTC_different_temperature} against Fig. 3(a) that the slight temperature differences have no noticeable impact on the heat transfer coefficients, thus justifying direct comparison of the data of Refs. \cite{Kloppstech2017,Messina2018} in Fig. 3(b).  

\newpage
\bibliography{SM}